# Novel two-dimensional Ca–Cl crystals with metallicity, piezoelectric effect and room-temperature ferromagnetism

Lei Zhang[1]*, Guosheng Shi[2]*, Bingquan Peng[1,3]*, Pengfei Gao[1]*, Liang Chen[4]*, Ni Zhong[5], Liuhua Mu[3], Han Han[3], Lijuan Zhang[3,6], Peng Zhang[1], Lu Gou[7], Yimin Zhao[1], Shanshan Liang[3], Jie Jiang[3], Zejun Zhang[3], Hongtao Ren[1], Xiaoling Lei[3,6], Long Yan[3#], Chungang Duan[5#], Shengli Zhang[1#] & Haiping Fang[3,6#]

[1]MOE Key Laboratory for Nonequilibrium Synthesis and Modulation of Condensed Matter, School of Science, Xi'an Jiaotong University, Xi'an 710049, China
[2]Shanghai Applied Radiation Institute, Shanghai University, Shanghai 200444, China
[3]Shanghai Institute of Applied Physics, Chinese Academy of Sciences, Shanghai 201800, China
[4]Zhejiang Provincial Key Laboratory of Chemical Utilization of Forestry Biomass, Zhejiang A&F University, Lin'an, Zhejiang 311300, China
[5]Key Laboratory of Polar Materials and Devices, Ministry of Education, East China Normal University, Shanghai 200241, China
[6]Zhangjiang Lab, Shanghai Advanced Research Institute, Chinese Academy of Sciences, Shanghai 201210, China
[7]Department of Applied Physics and State Key Laboratory for Manufacturing Systems Engineering, Xi'an Jiaotong University, Xi'an 710049, China

*These authors contributed equally to this work.
#Corresponding author. E-mail: fanghaiping@sinap.ac.cn (H.F.); zhangsl@xjtu.edu.cn (S.Z.); cgduan@clpm.ecnu.edu.cn (C.D.); yanlong@sinap.ac.cn (L.Y.)

**Recently we have reported the direct observation of two-dimensional (2D) Ca–Cl crystals on reduced graphene oxide (rGO) membranes, in which the calcium ions are only about monovalent (i.e. ~+1) and metallic rather than insulating properties are displayed by those CaCl crystals[1]. Here, we report the experimental observation and demonstration of the formation of graphene–Ca–Cl heterojunction owing to the metallicity of 2D Ca–Cl crystals, unexpected piezoelectric effect, room-temperature ferromagnetism, as well as the distinct hydrogen storage and release capability of the Ca–Cl crystals in rGO membranes. Theoretical studies show that the formation of those abnormal crystals is attributed to the strong cation–π interactions of the $Ca^{2+}$ with the aromatic rings in the graphitic surfaces. Since strong cation–π interactions also exist between other metal ions (such as $Mg^{2+}$, $Fe^{2+}$, $Co^{2+}$, $Cu^{2+}$, $Cd^{2+}$, $Cr^{2+}$ and $Pb^{2+}$) and graphitic surfaces[2], similar 2D crystals with abnormal valence state of the metal cations and corresponding abnormal properties as well as novel applications are highly expected. Those findings further show the realistically potential applications of such abnormal CaCl material with unusual electronic properties in designing novel transistors and magnetic devices, hydrogen storage, catalyzer, high-performance conducting electrodes and sensors, with a size down to atomic scale.**

Since the discovery of $C_{60}$[3], carbon nanotubes[4] and graphene[5], carbon continuously produces surprises. Up to now, most of the studies focus on the morphologies and properties of those carbon-based structures and their complexes with other components having normal stoichiometry. 3D and 2D Na–Cl and K–Cl systems with abnormal stoichiometries at extreme conditions, such as under high pressure[6] or on graphite/graphene[7] under ambient conditions have been reported in the last few years.

Recently, we have directly observed that the metallic 2D Ca–Cl crystals with about monovalent calcium ions on rGO membranes could be formed under ambient conditions[1]. Such metallic 2D Ca–Cl crystals are obtained by soaking rGO membranes in $CaCl_2$ solution below the saturated concentration under ambient conditions as described in our last work[1] (see PS1 of Supplementary Information, SI).

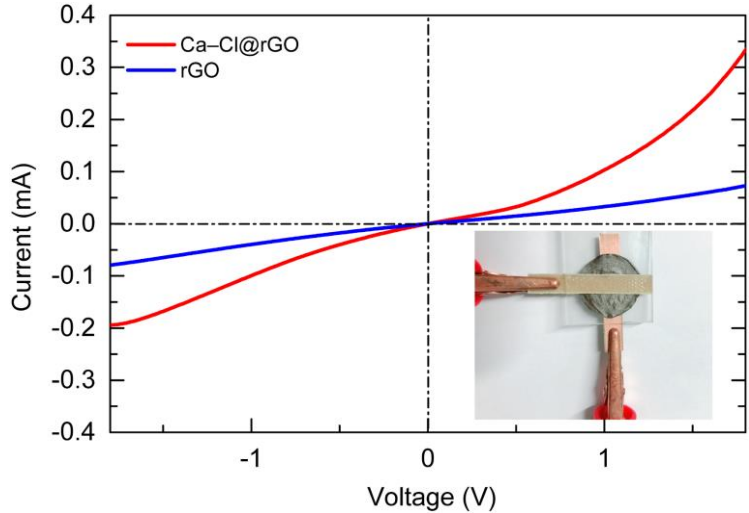

**Figure 1 | Heterojunction behavior of the dried Ca–Cl@rGO membrane.** The current–voltage plot shows a typical rectification behavior of the dried Ca–Cl@rGO membrane (red curve) compared with the dried rGO membrane (blue curve) under ambient conditions.

The metallic two-dimensional (2D) Ca–Cl crystals with monovalent calcium ions were proposed to have novel properties and greatly potential applications[1]. To demonstrate the existence of the "graphene–Ca–Cl" junction as theoretically suggested in our last work[1], we have measured the current–voltage response of the dried Ca–Cl@rGO junction membrane under positive and negative gate voltages. As a result, the current–voltage curve shows obviously asymmetric response, indicating that such junction indeed has typical rectification behavior (Fig. 1). Further, the ~+1 valence in these CaCl crystals provides distinct adsorption capacity with hydrogen storage, which not only can fill the insufficient adsorption of metallic calcium atoms but also overcome the exorbitant absorption of +2 calcium ions (Fig. 2, and see PS12 of SI for more information).

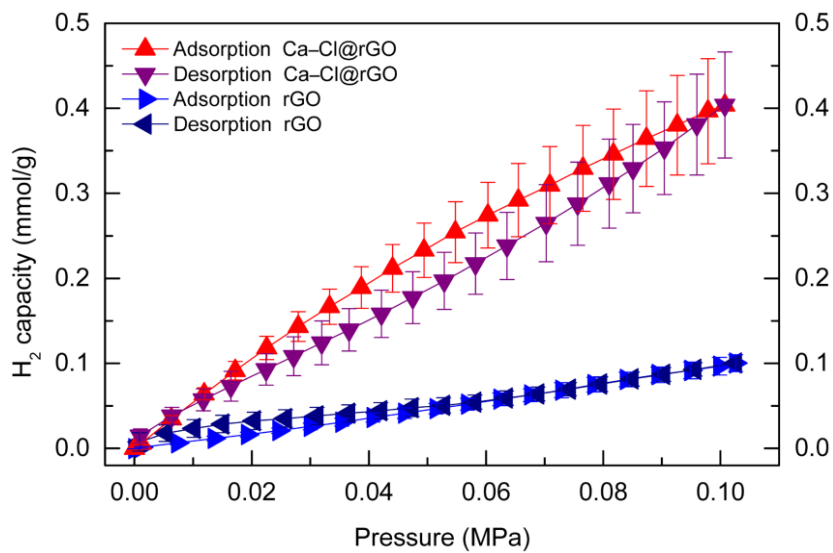

**Figure 2 | H$_2$ adsorption and desorption isotherms of the dried Ca–Cl@rGO and rGO membranes at 77 K.** The hydrogen adsorption and desorption curves of the rGO membrane are shown in blue and navy, respectively. The hydrogen adsorption and desorption curves of the dried Ca–Cl@rGO membrane are shown in red and purple, respectively.

Ca–Cl crystals in rGO membrane possess the unexpected piezoelectric property. DFT calculations show that, for 5% geometric elongation along the x direction, the charge distribution of the Ca, Cl and C atoms in model I has significant relocation (Fig. S30). Our experiment, where a dried Ca–Cl@rGO membrane was connected by two Cu foil electrodes (Figs. 3a and 3b), shows a maximal output voltage of ~4 mV when the membrane is bent with an angle of 90° (Fig. 3c), and the voltage decreases, reaches zero and may even overshoot to negative values when we remove the bending force. In contrast, we cannot detect a visible output voltage for the rGO membrane with the same bending angles. The voltage of the Ca–Cl@rGO membrane increases with increasing bending degree (Fig. 3d).

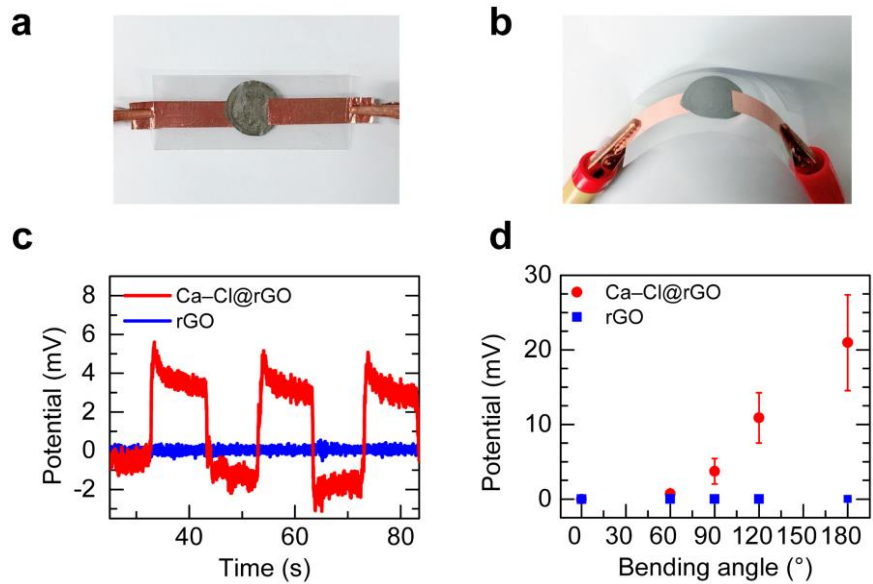

**Figure 3 | Piezoelectric effect of the dried Ca–Cl@rGO membrane. a**, Photo of the system in piezoelectricity measurement. A dried Ca–Cl@rGO or a dried rGO membrane is connected by two Cu foil electrodes fixed by flexible plastic plates. **b**, Photo of the system in piezoelectricity measurement of the dried membrane bent to a certain angle **c**, Typical voltage responses from a dried Ca–Cl@rGO membrane and a dried rGO membrane under periodic strain as a function of time for bending angles up to 90°. **d**, Piezoelectric outputs from a dried Ca–Cl@rGO membrane (red dots) and a dried rGO membrane (blue squares). The mean values and standard deviations of the output potentials were calculated from ten technical replicates.

Room-temperature ferromagnetism of the Ca–Cl crystals in rGO membrane was also experimentally observed and demonstrated. Fig. 4 shows that the saturation magnetic moment ($M_s$) of the dried Ca–Cl@rGO membrane is ~2.1 emu/g at 300 K, whereas $M_s$ = ~0.4 emu/g for the rGO membrane at the same temperature, showing a ~400% enhancement of magnetism by only introducing ~5% (wt) calcium to the membrane. Our calculations suggest that the possible origin of such strong ferromagnetism is the edge or defect effects of the CaCl crystals where there is an unpaired valence electron in $Ca^+$, and the saturation magnetic moment computed is consistent with the experimental observations (see PS11 of SI for more information).

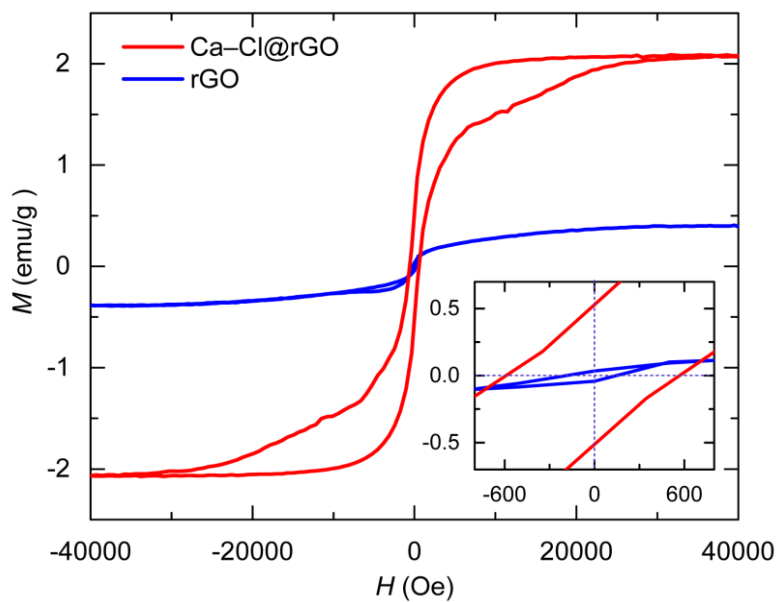

**Figure 4 | Magnetic moment *M* versus applied magnetic field *H* of the dried rGO and Ca–Cl@rGO membranes at 300 K after subtraction of the linear diamagnetic background.** The inset shows a low-field zoom of two curves from the main panel where excess moments and coercive force are seen clearly. The magnetic field is set to be perpendicular to the sample surface.

To summarize, based on the discovery of 2D CaCl crystals with unconventional stoichiometries[1], here we have experimentally demonstrated that the ~+1 valence of Ca ions possess various novel properties including metallic property and resulted graphene–Ca–Cl heterojunction, unexpected piezoelectric effect, room-temperature ferromagnetism, as well as the distinct hydrogen storage and release capability.

Such abnormal Ca–Cl crystals are attributed to the strong cation–π interactions between $Ca^{2+}$ and aromatic rings in the graphene surfaces. We expect similar abnormal crystals for other metal cations[2,7-14]. In fact under ambient conditions, Cu–Cl crystals with +1 copper ions can stably exist for several days (see PS7 of SI) and have room-temperature ferromagnetism (see PS11 of SI). This would help to overcome the difficulty of the short life time in the application as catalyzers for the +1 copper ion[15]. We note that it is highly expected that every metal element has room-temperature ferromagnetism via forming the correspondingly abnormal 2D crystals with an unpaired valence electron in the metal ions. Further, considering the wide distribution of metallic cations and carbon on earth, such nanoscale "special" compounds with previously unrecognized properties may be ubiquitous in nature.

**References and Notes:**

1   Zhang, L. *et al.* Two-dimensional Ca-Cl crystals under ambient conditions observed directly by cryo-electron microscopy. *arXiv:1812.07195* (2018).
2   Shi, G. *et al.* Ion enrichment on the hydrophobic carbon-based surface in aqueous salt solutions due to cation-π interactions. *Sci. Rep.* **3**, 3436 (2013).
3   Kroto, H. W., Allaf, A. W. & Balm, S. P. C60 - Buckminsterfullerene. *Chem. Rev.* **91**, 1213-1235 (1991).
4   Iijima, S. Helical Microtubules of Graphitic Carbon. *Nature* **354**, 56-58 (1991).
5   Novoselov, K. S. *et al.* Electric field effect in atomically thin carbon films. *Science* **306**, 666-669 (2004).
6   Zhang, W. W. *et al.* Unexpected Stable Stoichiometries of Sodium Chlorides. *Science* **342**, 1502-1505 (2013).
7   Shi, G. *et al.* Two-dimensional Na-Cl crystals of unconventional stoichiometries on graphene surface from dilute solution at ambient conditions. *Nat. Chem.* **10**, 776-779 (2018).
8   Chen, L. *et al.* Ion sieving in graphene oxide membranes via cationic control of interlayer spacing. *Nature* **550**, 415-418 (2017).
9   Gebbie, M. A. *et al.* Tuning underwater adhesion with cation-π interactions. *Nat. Chem.* **9**, 473-479 (2017).
10  Liu, J., Shi, G., Guo, P., Yang, J. & Fang, H. Blockage of water flow in carbon nanotubes by ions due to interactions between cations and aromatic rings. *Phys. Rev. Lett.* **115**, 164502 (2015).
11  Ma, J. C. & Dougherty, D. A. The cation-π interaction. *Chem. Rev.* **97**, 1303-1324 (1997).
12  Mahadevi, A. S. & Sastry, G. N. Cation-π interaction: its role and relevance in chemistry, biology, and material science. *Chem. Rev.* **113**, 2100-2138 (2013).
13  Shi, G. *et al.* Unexpectedly enhanced solubility of aromatic amino acids and peptides in an aqueous solution of divalent transition-metal cations. *Phys. Rev. Lett.* **117**, 238102 (2016).
14  Xiu, X., Puskar, N. L., Shanata, J. A., Lester, H. A. & Dougherty, D. A. Nicotine binding to brain receptors requires a strong cation-π interaction. *Nature* **458**, 534-537 (2009).
15  Gawande, M. B. *et al.* Cu and Cu-Based Nanoparticles: Synthesis and Applications in Review Catalysis. *Chem. Rev.* **116**, 3722-3811 (2016).

**Supplementary Information:**
Materials and Methods
Supplementary Text
Figs. S1 to S40
Tables S1 to S6
References
Caption for Movie S1